\documentclass[twocolumn,showpacs,preprintnumbers,amsmath,amssymb]{revtex4}

\usepackage{graphicx}
\usepackage{dcolumn}
\usepackage{bm}

\begin{document}

\preprint{APS/123-QED}

\title{Modeling Time Series of Real Systems using Genetic Programming}

\author{Dilip P. Ahalpara}
 \email{dilip@ipr.res.in}
 \affiliation{Institute for Plasma Research, Near Indira Bridge, Gandhinagar-382428, India}
\author{Jitendra C. Parikh}%
 \email{parikh@prl.res.in}
\affiliation{Physical Research Laboratory, Navrangpura, Ahmedabad-380009, India}

\date{\today}

\begin{abstract}
Analytic models of two computer generated time series (Logistic map and Rossler system) and two real time series (ion saturation current in Aditya Tokamak plasma and NASDAQ composite index) are constructed using Genetic Programming (GP) framework. In each case, the optimal map that results from fitting part of the data set also provides a very good description of rest of the data. Predictions made using the map iteratively range from being very good to fair. 
\end{abstract}

\maketitle

\section{\label{sec:level1}Introduction}
The problem of constructing models of complex dynamic systems from its time series is of great interest in various fields of science and economics. This is important both for insights they may provide into the dynamics and for making predictions. Since the dynamics of such systems is expected to be non linear in nature, different methods have been suggested for constructing the models \cite{TS:Weigend}, \cite{TS:Kantz}. These include local linear models, radial basis function approach, artificial neural networks (ANN), genetic algorithm (GA) and genetic programming (GP). In this note, we work with the Genetic Programming framework similar to that used by Szpiro \cite{GP:Szpiro} and extend it in some ways. A brief description of GP is given in Sec. II.

The approach is first tested with two deterministic chaotic time series (Logistic map and Rossler system) \cite{Chaos:Strogatz}. We obtain an excellent representation of the data in terms of a non-linear regressive map. In order to further validate the model we also make dynamic predictions. In our view, this is a crucial test of the model. For the Logistic map we find impressive results but for the Rossler map the results are not so good (Sec. III).

We next consider two measured or observed time series. The first one is the time series of measurement of ion saturation current in Aditya Tokamak plasma \cite{Aditya:Jha} and the second one is the finanical NASDAQ composite index. These series are non-stationary and exhibit fluctuations that are statistical in nature. In view of this we first separate the fluctuations from the underlying mean behaviour of the series using wavelet transforms \cite{Wavelet:Daub} A brief account of wavelets is given in Sec. IV A. Further, as in statistical physics (e.g. Langevian equation for Brownian particle) we assume mean dynamics to be deterministic and fluctuations to be stochastic. In order to model the mean dynamics in the GP framework we first construct an appropriate embedding in the reconstructed phase space for each time series. This is briefly summarized in Sec. IV B. The properties of the embedding together with genetic programming then gives us an analytical expression for the map. Interestingly, the optimal non-linear model that is generated in this manner is of the Pade form \cite{Pade:NR}. They have many interesting mathematical properties, e.g. same or better convergence properties as compared to power series and can model functions with singularities. We find that the maps describe the dynamics quite well and give good one step predictions. However, when iterated, the prediction deteriorates rapidly. Our results for the Aditya Tokamak data and those for the NASDAQ composite index are presented in Sec. IV C and D respectively. Sec. V contains some concluding remarks.

\section{Genetic Programming}
Under the umbrella of Evolutionary Algorithms (EA) \cite{GA:Holl}, \cite{GA:Gold}, \cite{GA:Fog}, various approaches like Genetic Algorithm (GA), Genetic Programming (GP), Evolution Strategies (ES) etc have been framed that are aimed at solving complex search and optimization problems. The approaches differ from one another by the implementation details of genetic structures and the use of various genetic operators. In the present note we have used Genetic Programming (GP) framework that uses a nonlinear structure for chromosomes (details given in Appendix) that represent candidate solutions, as against the more popular approach of Genetic Algorithm (GA) \cite{GA:Melanie} that uses linear structures of chromosomes.

Genetic Programming (GP) uses an iterative computation to progressively get better and better candidate solutions. We first initialize the population P(time t=0) randomly with chromosomes generated as per a template structure, such as

$((A \otimes B) \otimes (C \otimes D))$

where A, B, C, D are either time lagged variables (to be described later) or real numbers and $\otimes$ is one of the arithmatic operators +, -, $\times$ or $\div$. The population P(t) is then iterated by following the steps given below:
\begin{enumerate}
  \item Evaluate chromosomes in P(t) using an objective function that is a measure of fitness. Sort the population P(t) according to fitness values.
  \item Preserve a portion of good chromosomes of P(t) by copying them on another population P(t+1) using the copy operator. This would assure that the best chromosomes found so far are not lost due to the application of genetic operations on P(t). This feature is known as {\it elitism}.
  \item Select pairs of chromosomes from the remaining portion of P(t) (typically using a roulette wheel selection criteria that gives more preference to chromosomes having higher fitness values) and recombine the pairs (i.e. parents) stochastically to generate offsprings and put them in P(t+1).
  \item Mutate offsprings in P(t+1) stochastically.
  \item Steps 3 and 4 give rise to the next generation population P(t+1).
  \item Replace newly generated population P(t+1) with P(t)
  \item Advance time t to $t+1$.
  \item Verify the termination criteria (i.e. whether generation number t has crossed a preassigned upper limit or whether convergence for fitness values of top chromosomes have been achieved).
\end{enumerate}
Once satisfactory solution(s) have been found, the iteration is stopped.

We have followed the general outline of Genetic Programming (GP) as in Szpiro \cite{GP:Szpiro} to fit a given data set of time series. For the time series considered presently, we use 500 points for fitting the data and then carry out an out-of-sample prediction. The prediction is done using a one-step approach and a dynamic iterative approach, to be described later.

For all the time series considered presently, we assume the map equation to have the form
\begin{equation}
X_{t} = f(X_{t-\tau}, X_{t-2\tau}, X_{t-3\tau}, ... X_{t-d\tau})
\label{eq:MapEquation}
\end{equation}

where f represents a function involving time series values $X_{t}$ in the immediate past, arithmatic operators (+, -, $\star$ and $\div$) and numbers bound between -10 and 10 with a precision of 1 digit; d represents number of previous time series values that may appear in the function and $\tau$ represents a time delay (to be described later in Sec. IV B).

The sum of squared errors,
\begin{equation}
\bigtriangleup^{2} = \sum_{i=1}^{i=N} (X_{i}^{calc} - X_{i}^{given})^{2}
\label{eq:SumSqErrors}
\end{equation}

is minimized, where N represents number of $X_{t}$ values [Eq. (\ref{eq:MapEquation})] that are fitted during the GP optimization.

For a given chromosome, the lower the above sum of squared errors, the fitter is the chromosome. The fitness measure derived from $\bigtriangleup^{2}$ is defined as:
\begin{equation}
R^{2} = 1 - \frac{\bigtriangleup^{2}}{\displaystyle\sum_{i=1}^{i=N} (X_{i}^{given} - \overline{X_{i}^{given}})^{2}}
\label{eq:RSquare}
\end{equation}

where $\overline{X_{i}^{given}}$ is the average of all $X_{t}$ (Eq. \ref{eq:MapEquation}) to be fitted.

As described in \cite{GP:Szpiro}, the Genetic programming is discouraged to overfit by generating longer strings of chromosomes. This is achieved by modifying the fitness measure as follows,
\begin{equation}
r = 1 - (1 - R^{2})\frac{N-1}{N-k}
\label{eq:RSquareModified}
\end{equation}

where N is the number of equations to be fitted in the training set and k is the total number of time lagged variables of the form $X_{t-\tau}$, $X_{t-2\tau}$, ... etc (including repetitions) occurring in the given chromosome. This modified fitness measure prefers a parsimonious model. For $R^{2}$ close to 0, r can be negative.

\section{GP Model for time series of known systems}
We consider the time series of 2 known systems, namely 1) Logistic map and 2) Rossler system.

\subsection{Time Series of Logistic Map}
The Logistic map is defined by the equation
\begin{equation}
X_{n+1} = r X_{n} (1-X_{n})
\label{eqn:LogisticMap}
\end{equation}

We have chosen the control parameter r as 3.891 so that Eq. (\ref{eqn:LogisticMap}) generates a chaotic time series. Choosing $X_{0}$=0.1 and bypassing initial 2000 transient points we generate the time series.

We then use GP to fit the data set of 500 points to get the best possible map function using d=1 and $\tau$=1. The fit obtained is very good giving $\bigtriangleup^{2}$=3.523*$10^{-10}$ and modified fitness r=1.0. The map equation for $X_{t}$ as a function of time lagged variables as obtained by GP is as follows:
\begin{tiny}
\begin{eqnarray}
X_{t}&=&X_{t}^{N} /  X_{t}^{D}                                                  \nonumber \\
X_{t}^{N}&=&-3.9(X_{t1}+2.9)X_{t1}(X_{t1}-0.0521)(X_{t1}-0.0693)(X_{t1}-0.976)  \nonumber \\
&&(X_{t1}-1.0)^{2}(X_{t1}-7.46)(X_{t1}^{2}+2.25X_{t1}+1.46)                     \nonumber \\
&&(X_{t1}^{2}+1.03X_{t1}+1.63)(X_{t1}^{2}-1.41X_{t1}+1.82)(X_{t1}^{2}-3.49X_{t1}+3.30)          \nonumber \\
X_{t}^{D}&=&(X_{t1}+2.9)(X_{t1}-0.0522)(X_{t1}-0.0693)(X_{t1}-0.976)            \nonumber \\
&&(X_{t1}-1.0)(X_{t1}-7.46)(X_{t1}^{2}+2.26X_{t1}+1.46)(X_{t1}^{2}+1.03X_{t1}+1.63)                    \nonumber \\
&&(X_{t1}^{2}-1.41X_{t1}+1.83)(X_{t1}^{2}-3.49X_{t1}+3.3)
\label{eqn:LogiMap}
\end{eqnarray}
\end{tiny}

We use the notation $X_{tm}$=$X_{t-m*\tau}$ and show double precision numbers to only 3 significant digits for the sake of simplicity. It can be seen that many of the factors in the numerator and the denominator approximately cancel (apart from higher precision effect) and the resulting simplified form of the map is $X_{t}$$\approx3.9 X_{t-\tau}(1-X_{t-\tau}$) that is remarkably close to the actual map equation.

The normalized mean square error (NMSE),
\begin{equation}
NMSE = \frac{1}{N}  \frac{\displaystyle\sum^N_{i=1} [X_i^{calc} - X_i^{given}]^{2}  }
{variance \ of \ N \ data \ points}
\label{eqn:NMSE}
\end{equation}

is used as an index for the goodness of fit. For the above GP fit for the dataset of 500 points of Logistic map, we get NMSE=9.84294*$10^{-12}$.

The map is then used to make an out-of-sample one-step prediction (in which given time lagged values are successively used to predict the next dataset value) starting at different regions in the time series.

\begin{figure}
\centering
    {\resizebox{!}{6.5cm}{%
       \includegraphics{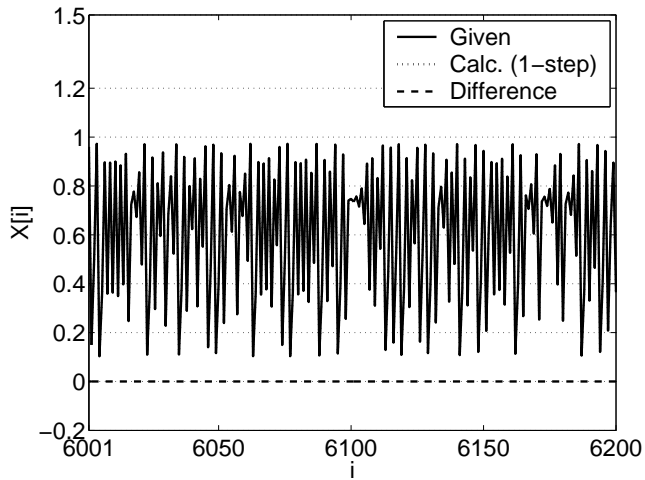}}}
\vspace{-0.1in}
\caption{\label{fig:LogisticPredict6001}Out-of-sample one-step prediction of 200 points for Logistic time series beginning at data point 6001.}
\end{figure}

\begin{table}
\caption{\label{tab:LogisticNMSE}NMSE (Eq. \ref{eqn:NMSE}) for Logistic time series using one-step prediction of 200 points starting at different data points.}
\begin{ruledtabular}
\begin{tabular}{rl}
Starting data point&NMSE                    \\                    \hline
 1001&8.18436e-12                           \\
 2001&1.01925e-11                           \\
 4001&8.78014e-12                           \\
 6001&9.97224e-12                           \\
 8001&8.80229e-12                           \\
\end{tabular}
\end{ruledtabular}
\end{table}

Fig. \ref{fig:LogisticPredict6001} shows out-of-sample one-step prediction starting at data point 6001. Table \ref{tab:LogisticNMSE} shows NMSE values for 200 point one-step prediction starting at different data points outside the fitted dataset. Note that the predictions are in almost perfect agreement with the data.

\begin{figure}
{\resizebox{!}{6.5cm}{%
  \includegraphics{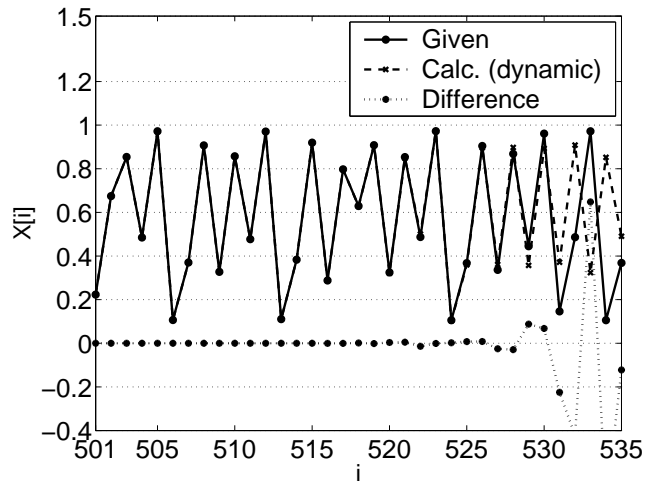}}}
\caption{\label{fig:LogiPredict501} Out of sample prediction for Logistic map using GP solution.}
\end{figure}

As for any other model, the real test of the GP solution lies in making a dynamic prediction, in which it is assumed that data beyond the fitted dataset are not available and hence calculated values are progressively used beyond the fitted dataset. Fig. \ref{fig:LogiPredict501} shows the dynamic prediction for 35 points beyond the dataset of 500 fitted points.

It is seen that the dynamic prediction holds good for around 28 steps (the \% error at point no. 28 being 1.48*$10^{-4}$). After that the prediction deteriorates due to the chaotic nature of the time series. As is known, Lyapunov exponents provide a quantitative measure of the sensitivity of the initial condition in a given chaotic system (i.e. the measure of the divergence of neighboring trajetories exponentially in time). It is therefore useful to calculate the Lyapunov exponent for the Logistic map. We have used the programs given in Kantz et al \cite{TS:Kantz} for the calculation of Lyapunov exponents. The Lyapunov exponent for the Logistic time series considered is 0.471. It may be noted that this method for the calculation of Lyapunov exponent has a possible element of small error due to selection of linear part from multiple curves for finding its slope. Using this Lyapunov exponent, an initial error of 1.61*$10^{-7}$ in the first step of dynamic prediction is expected to grow to around 0.1 (and rapidly to higher values there after) around 28 points. Thus in this case we understand the inherent limitation of iterative dynamic prediction due to chaotic nature of the time series.

\subsection{Time Series of Rossler system}
Next we consider the time series generated from discretized Rossler equations \cite{GP:Szpiro}:
\begin{eqnarray}
X_{t+\delta} &=& X_{t} - [(Y_{t}+Z_{t})]\delta              \notag   \\
Y_{t+\delta} &=& Y_{t} + [(X_{t}+aY_{t})]\delta             \notag   \\
Z_{t+\delta} &=& Z_{t} + [b+X_{t}Z_{t}-cZ_{t}]\delta
\label{eq:Rossler}.
\end{eqnarray}

The parameters in Rossler equations  are selected as in \cite{GP:Szpiro} and are: a=0.2, b=0.2 and c=5.7. The time series for $X_{t}$, $Y_{t}$ and $Z_{t}$ are generated with initial conditions as $x_{0}$=-1, $y_{0}$=0 and $z_{0}$=0 and $\delta$=0.02 and using every $50^{th}$ point of the series generated.

A GP fit is then made on the dataset of 500 points with values d=8 and $\tau$=1 used by Szipro \cite{GP:Szpiro}. The map equation generated from the fit using GP gives fitness value as 0.9874 and the map equation is:
\begin{tiny}
\begin{eqnarray}
X_{t}&=&X_{t}^{N} /  X_{t}^{D}                                                  \nonumber \\
X_{t}^{N}&=&(0.111X_{t5}+1.257)(X_{t1}-X_{t2}-0.14(63.83-15.1X_{t4}   \notag \\
&&+(X_{t1}+2X_{t4}+10.175)(0.149(X_{t4}+10)X_{t5}(X_{t1}-X_{t2}-1.0)+1.061)  \notag \\
&&-0.139X_{t6}(X_{t1}+2X_{t4}+10.175)(3X_{t2}+0.175)+X_{t2}+X_{t7}           \notag \\
&&+\frac{0.139(210.69-7.525X_{t1}-\frac{76.567}{X_{t2}}-7.9X_{t4})}{481.08-X_{t6}+X_{t5}+59.04X_{t4}+X_{t1}})                                                                    \notag \\
X_{t}^{D}&=&(X_{t1}+7.2X_{t4}-X_{t5}-X_{t6}+57)
\label{GPSoln:Rossler}.
\end{eqnarray}
\end{tiny}

We use the notation $X_{tm}$=$X_{t-m*\tau}$ and show double precision numbers to only 3 significant digits for the sake of simplicity.

\begin{figure*}
\centering
  \begin{tabular}{cc}
    {\resizebox{!}{6.5cm}{%
       \includegraphics{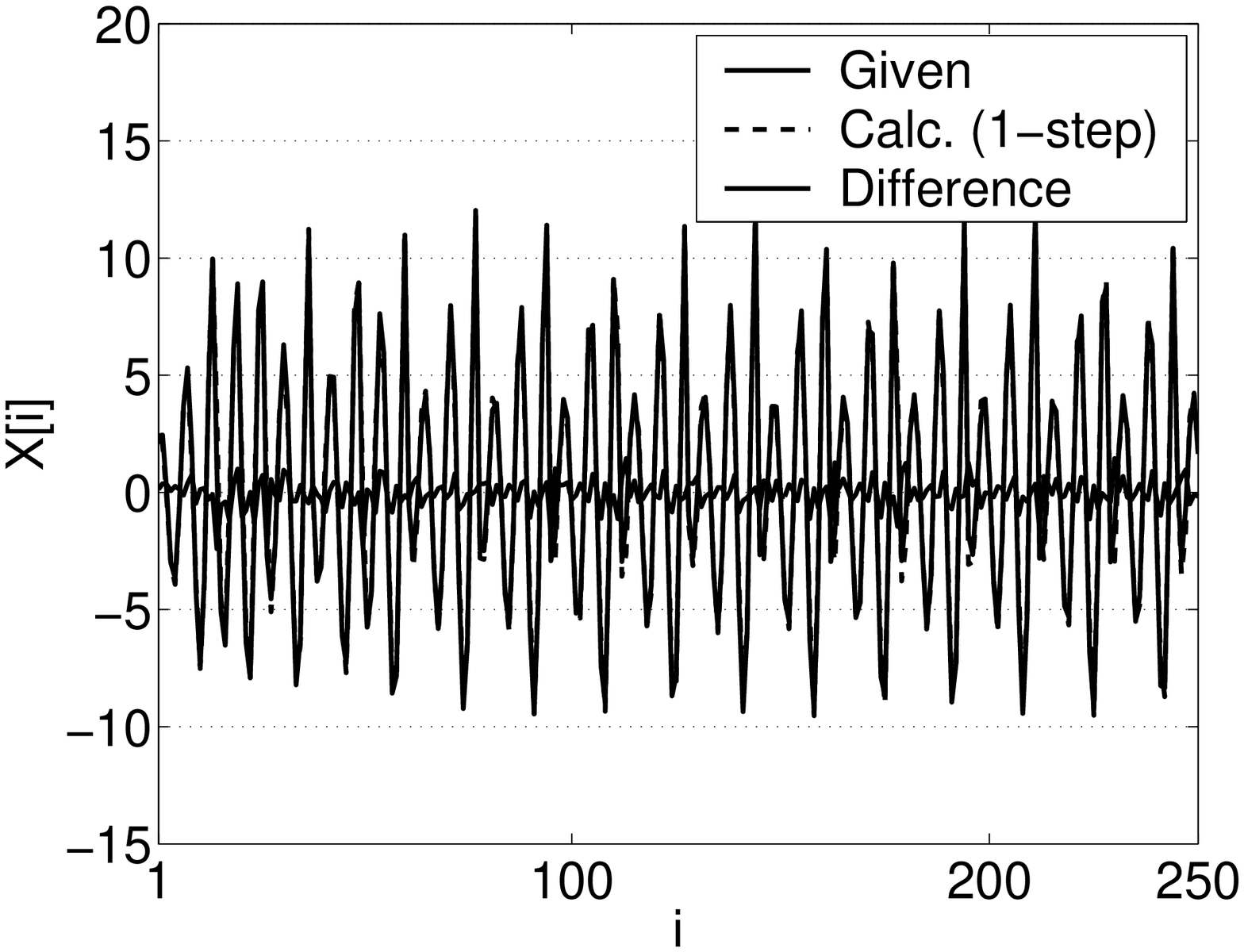}}}
    {\resizebox{!}{6.5cm}{%
       \includegraphics{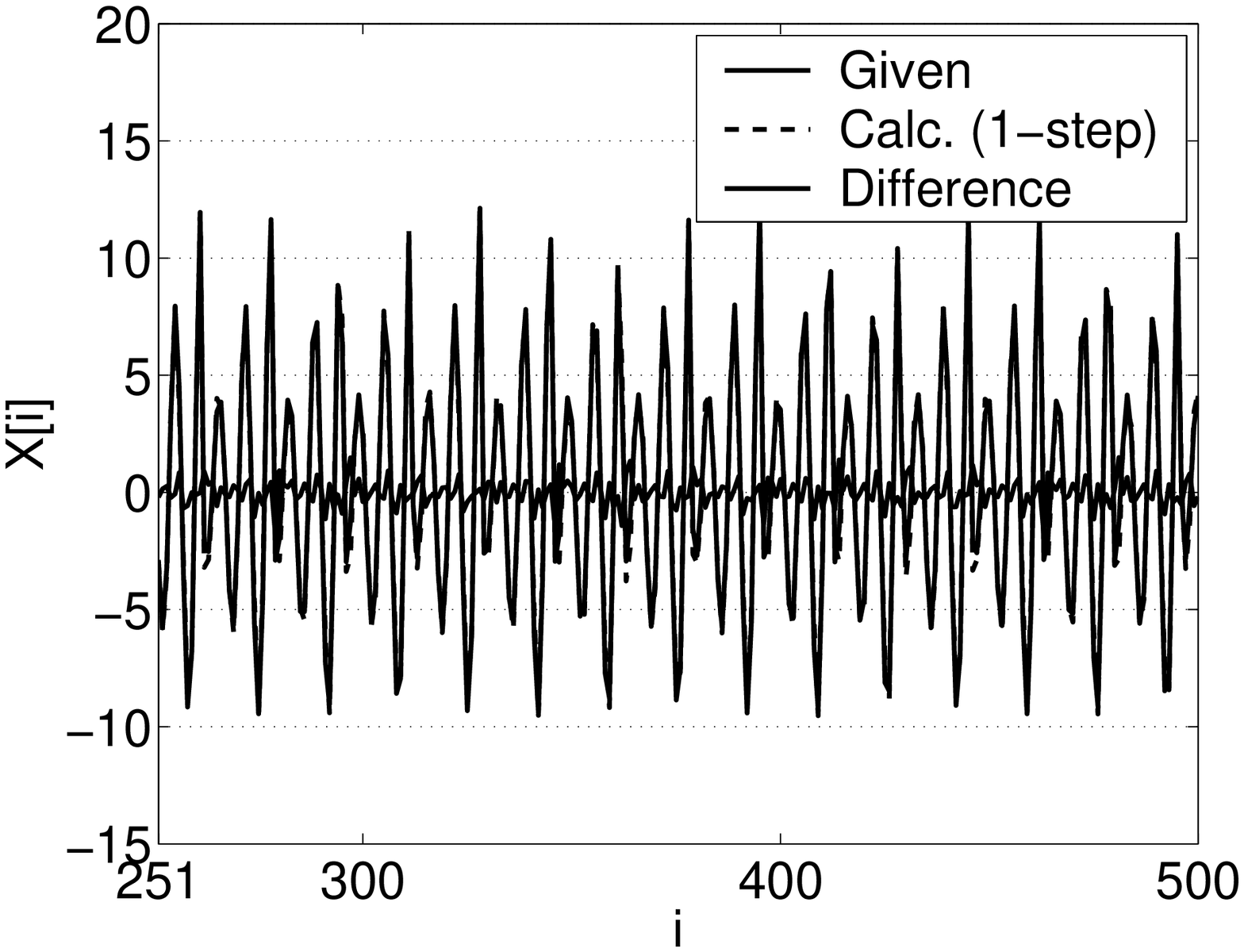}}}
  \end{tabular}
  \vspace{-0.15in}
    \begin{verbatim}                             (a)                                                (b)\end{verbatim}
\vspace{-0.1in}
\caption{\label{fig:RossFit500}GP fit of 500 points for Rossler time series. (a) and (b) show initial and last 250 points of the 500 point time series respectively.}
\end{figure*}

Fig. \ref{fig:RossFit500} shows comparision between given and GP calculated values of 500 data points. The thick line close to 0.0 marks the difference between the given and calculated values and indicates that the fit is quite good.

Next we carry out an out-of-sample prediction beyond the fitted dataset of 500 points. 

\begin{figure*}
\centering
  \begin{tabular}{cc}
    {\resizebox{!}{6.5cm}{%
       \includegraphics{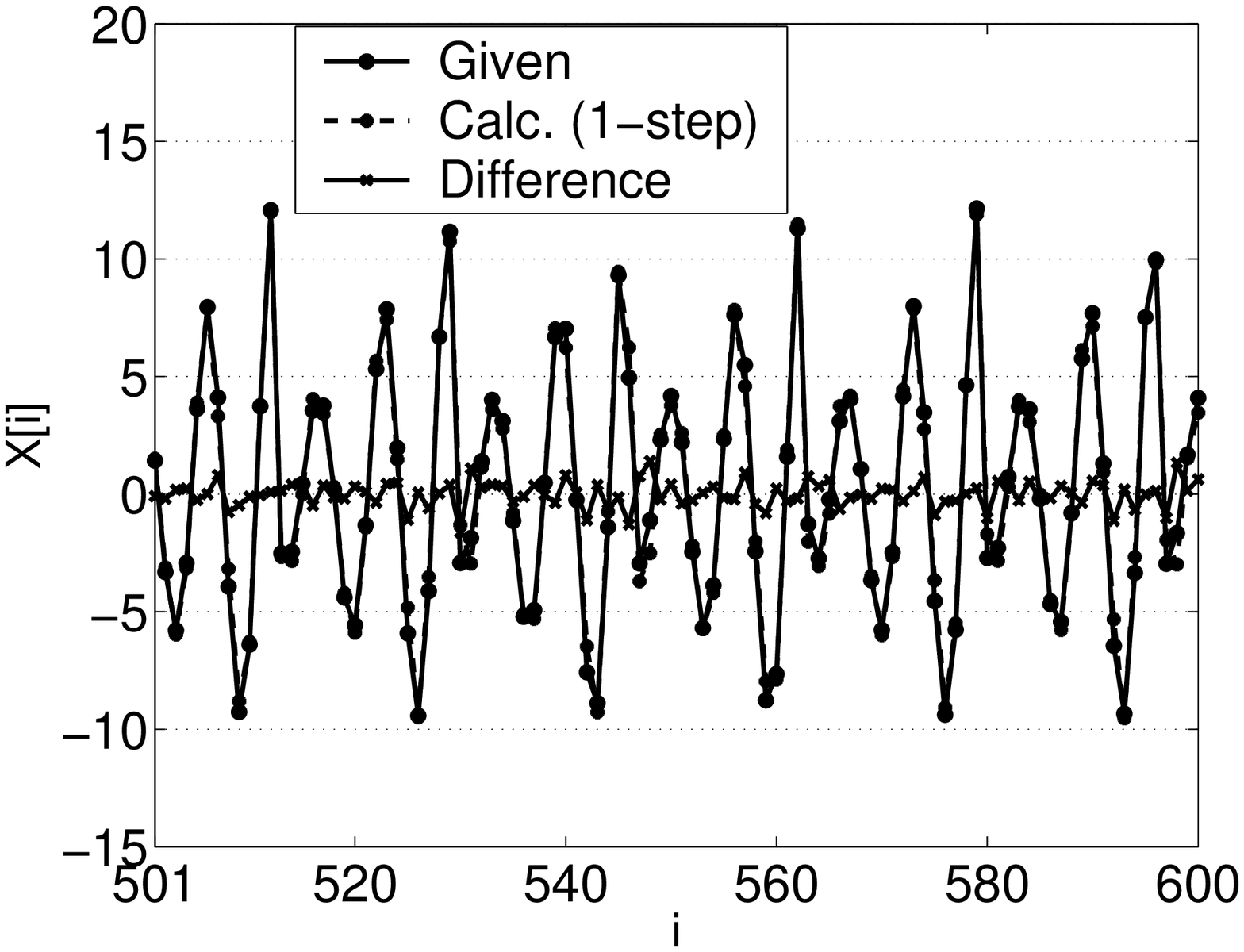}}}
    {\resizebox{!}{6.5cm}{%
       \includegraphics{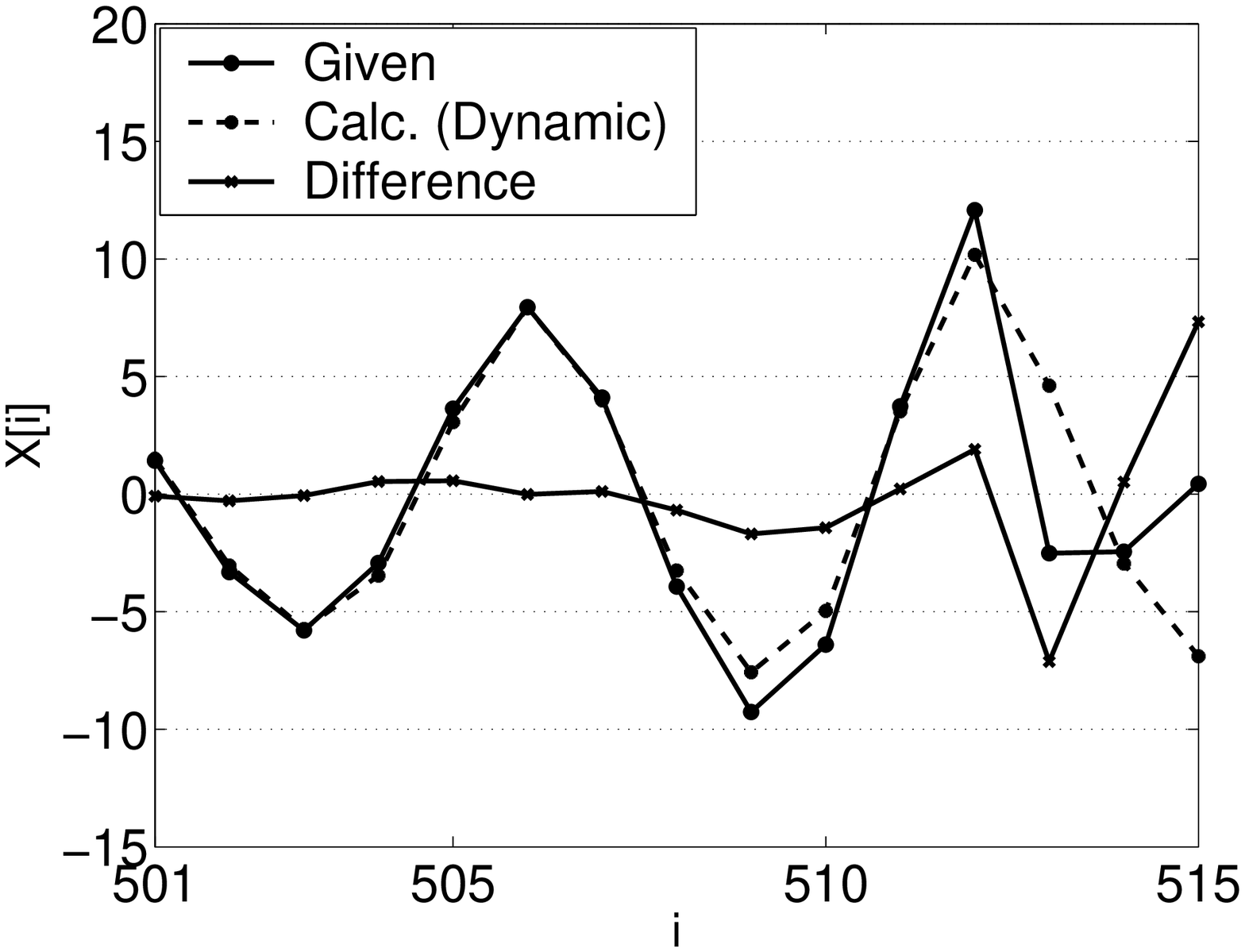}}}
  \end{tabular}
  \vspace{-0.15in}
    \begin{verbatim}                             (a)                                                (b)\end{verbatim}\vspace{-0.1in}
\caption{\label{fig:RossPredict501}Out of sample prediction beyond the fitted dataset of 500 points for Rossler time series. (a) and (b) show 1-step prediction (for 100 points) and dynamic prediction (for 15 points) respectively.}
\end{figure*}

Fig. \ref{fig:RossPredict501} (a) shows 1-step prediction for 100 points. It can be seen that the 1-step prediction is quite good. The NMSE (Eq. \ref{eqn:NMSE}) value for the one-step fit is 0.0118. Next a dynamic prediction is made using the GP solution. Fig. \ref{fig:RossPredict501} (b) shows comparison between actual and predicted values. As can be seen from the figure, the prediction is quite good for around 12 points beginning from data point 501. The Lyapunov exponent for the time series considered is 0.22. Using this value of Lyapunov exponent, it is estimated that the initial error of 0.08438 would grow to 1.1825 on the $12^{th}$ step and to higher values rapidly thereafter. This again is in close agreement to the trend observed in Fig. \ref{fig:RossPredict501} (b) showing the dynamic prediction by GP solution.

The NMSE value for dynamic prediction of 12 points beyond the fitted dataset is 0.048.

\section{Analysis of time series of real systems}
Next we consider the time series of 2 real systems, namely 1) ion saturation current in Aditya Tokamak plasma and 2) financial NASDAQ composite index. Since these series have stochastic fluctuations, it is necessary to filter them to obtain the trend. The trend is then modelled using GP approach. It is therefore required to filter these series using an appropriate method.

\subsection{Smoothening the Time Series Data Set for real systems}
A number of methods exist in the literature and have been used for separating fluctuations from trend in a time series. In this context, it is important to point out that most time series of real systems with complex dynamics are non-stationary in nature. Consequently in such cases, we need to employ a suitable method to separate the fluctuations from the trend. We use discrete wavelet transforms (DWT) \cite{Wavelet:Daub} for this purpose. Such an approach has recently been proposed by Manimaran et al \cite{TS:Mani} and Ahalpara et al \cite{TS:DPAJCP}. The basic reason for our choice of DWT is related to their nice mathematical properties. In the present context, it is sufficient to note that (i) DWT provides a complete orthonormal basis to decompose a non-stationary  signal and (ii) wavelet functions have a finite number of moments that are zero. In our work we have used length-4 Daubechies-4 (Db-4) wavelet transform. It is one of the simplest and smallest (even) length wavelet transform that is smooth. We next describe in brief our wavelet based procedure.

Given a time series composed of n points, namely $X_i$, i=1, 2, ... n, we first carry out a forward Db-4 wavelet transformation \cite{Wavelet:Daub} that gives n wavelet coefficients. Of these, half (n/2) are low pass coefficients that describe the average behavior locally and the other half (n/2) are high pass coefficients corresponding to local fluctuations. In order to obtain a smooth time series, the high pass coefficients are set to zero and then an inverse Db-4 transformation is carried out. This results in smoothening of the data set (see Manimaran et al \cite{TS:Mani} and Ahalpara et al \cite{TS:DPAJCP}). While using the Db-4 transform it has been observed that due to fixed boundary of the data set, rapid fluctuations are observed towards the beginning and the end of the smoothened data set. In order to remove this spurious effect, we do a padding of the data set by adding constant valued n/2 data points at the beginning and n/2 data points at the end of the time series. The constant value matches with the value of the first and the last data point respectively. The forward Db-4 transformation is then applied on the data set having 2n data points. Having smoothened the data set by an inverse Db-4 transformation, the padded data sets (containing the spurious effect) are removed thereby getting an improved smoothened time series of the n data set points. We thus obtain a {\em level one} time series of trends in which fluctuations at the smallest scale have been filtered out and the trend extracted after applying Db-4 transform. One can repeat the entire process on {\em level one} series to filter out fluctuations at the next higher time scale to get a {\em level two} smoothened series and so on. It is worth emphasizing that the low-pass wavelet coefficients give a representation in the transformed space of the smooth determistic part of the time series.

\subsection{State Space Reconstruction}
We use the standard time delay embedding approach using the smoothened time series as a means of reconstructing the vector space that is equivalent to the original state space of the system. In order to carry out the embedding we need to determine 2 parameters, namely time delay $\tau$ and embedding dimension d.

Average mutual information analysis is used to obtain the time delay $\tau$ and the number of false nearest neighbors analysis is used to obtain the embedding dimension d. These two methods are described by Abarbanel et al \cite{TS:Abarbanel}.

We use the prescription I($\tau)$/I(0) $\approx$ 0.2 suggested by Abarbanel et al \cite{TS:Abarbanel} for choosing the time delay $\tau$.  Here $I(\tau)$ represents average mutual information as a function of time lag $\tau$.

The dimension d is fixed by choosing the smallest dimension for which number of false neighbors become zero. Further we require that the number of false neighbors consistently remains zero thereafter for higher dimensions. We have used this criterion for all the time series considered in present analysis.

\subsection{Aditya Tokamak data}
We first consider the experimental time series of ion saturation current in Aditya Tokamak plasma \cite{Aditya:Jha}. This series is first smoothened using Db-4 transformation with level=1, 2 and 3. 

The time lag $\tau$ and dimension d of the embedded vectors are found by average mutual information analysis and number of false neighbors analysis. These are shown in Table \ref{tab:AdityaTauDim}.

\begin{table}
\caption{\label{tab:AdityaTauDim}Time lag $\tau$ and dimension d obtained for Aditya time series using average mutual information and \% of false neighbors analysis.}
\begin{ruledtabular}
\begin{tabular}{ccc}
Level&Time lag $\tau$&Dimension d        \\  \hline
 1&1&5                                   \\
 2&1&5                                   \\
 3&1&8                                   \\
\end{tabular}
\end{ruledtabular}
\end{table}

A GP fit is made on the datasets of Aditya time series having 500 points each with Db-4 level 1, 2 and 3 time series. As shown in Table \ref{tab:AdityaGPFitness}, the fitness value for the fits for these series are comparable and is maximum for level 3 series. It may be noted that the original Aditya time series is rather coarsely measured and therefore we find it more appropriate to use Db-4 smoothened level 2 and 3 series and not consider level 1 series. It also turns out that the GP solution for Db-4 level 1 Aditya time series is quite involved one and is also therefore of less interest to analyse further.

\begin{table}
\caption{\label{tab:AdityaGPFitness}The fitness parameters for Db-4 smoothened level 1, 2 and 3 Aditya time series obtained using GP fit on datasets of 500 points each}
\begin{ruledtabular}
\begin{tabular}{ccc}
 &$\bigtriangleup^{2}$&Fitness                  \\                    \hline
Aditya series (Db-4 level=1)&3.3685&0.9724      \\
Aditya series (Db-4 level=2)&2.6329&0.9776      \\
Aditya series (Db-4 level=3)&1.8594&0.9834      \\
\end{tabular}
\end{ruledtabular}
\end{table}

We have therefore considered here the analysis for Db-4 level 2 and 3 Aditya time series.  

The best solutions found by GP, having fitness value of 0.9776 (for level=2) and 0.9834 (for level=3) Aditya time series are:
\begin{widetext}
\begin{eqnarray}
X_{t}^{level=2}&=&X_{t1}+\frac{X_{t1}-X_{t2}}{X_{t1}+\frac{X_{t1}+X_{t3}-X_{t1}^{2}X_{t3}}{X_{t2}(\frac{X_{t1}X_{t3}+X_{t1}X_{t4}-X_{t5}^{2}}{X_{t5}})(X_{t3}+\frac{X_{t1}}{X_{t1}+X_{t2}-X_{t4}+X_{t5}+\frac{X_{t2}}{3X_{t1}+X_{t3}-X_{t1}^{2}X_{t3}+X_{t5}}})}}  \nonumber \\
X_{t}^{level=3}&=&\frac{X_{t1}^{2}}{X_{t2}+\frac{0.336X_{t1}^{2}}{\frac{3X_{t2}}{2}+\frac{X_{t2}-\frac{7.6X_{t1}}{7.8X_{t1}+210X_{t1}X_{t4}-1107.4}}{2(2.12X_{t1}+0.11X_{t1}^{2}-\frac{0.84}{X_{t1}X_{t2}}+\frac{0.23X_{t1}}{2.1X_{t1}-X_{t1}^{2}+X_{t2}})}}} \label{eq:GPSoln_Aditya}.
\end{eqnarray}
\end{widetext}

We use the notation $X_{tm}$=$X_{t-m*\tau}$ and show double precision numbers to only 3 significant digits for the sake of simplicity.

\begin{figure}
\centering
  \begin{tabular}{cc}
    {\resizebox{!}{3.4cm}{%
       \includegraphics{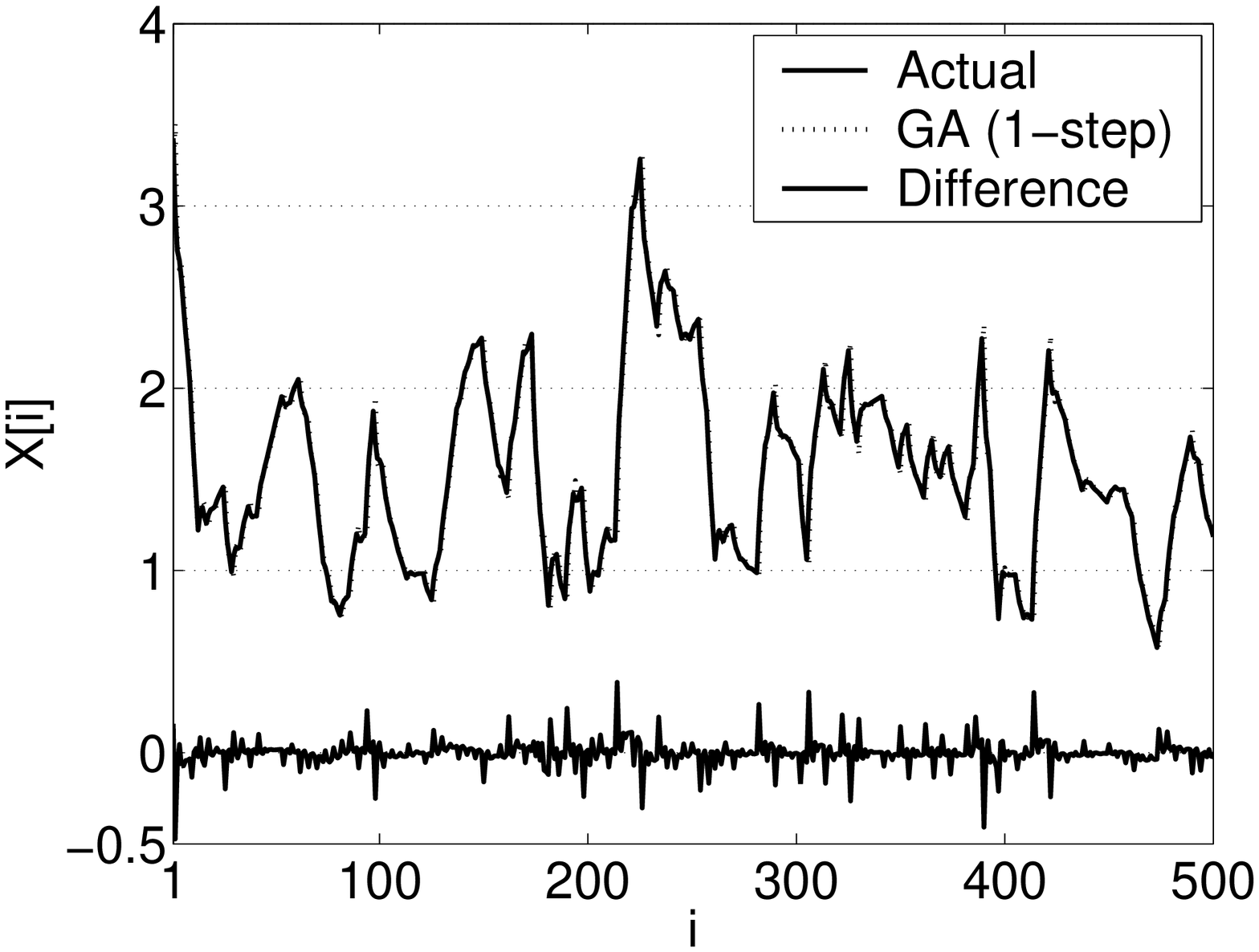}}}
    {\resizebox{!}{3.4cm}{%
       \includegraphics{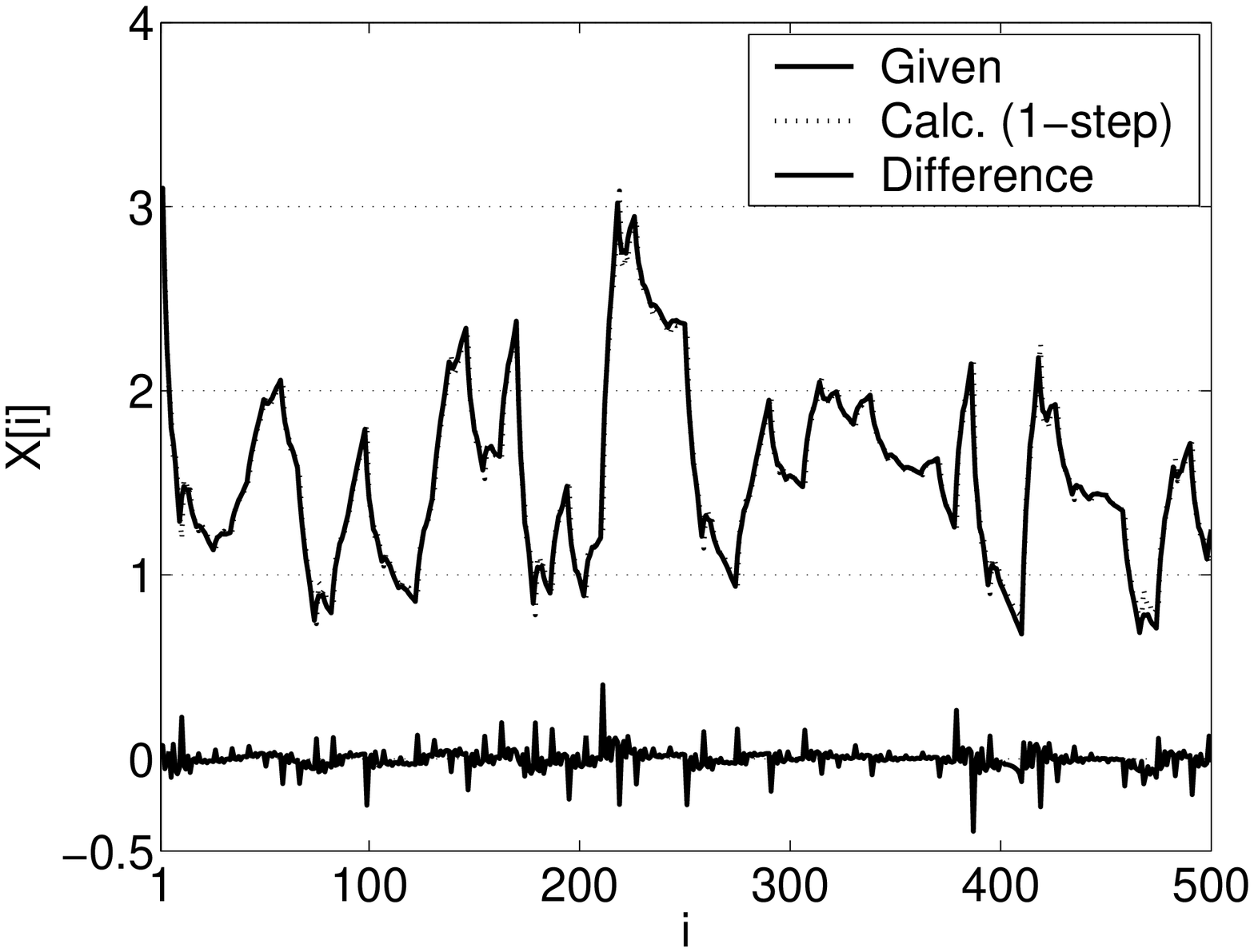}}}
  \end{tabular}
  \vspace{-0.15in}
    \begin{verbatim}              (a)                   (b)\end{verbatim}
\vspace{-0.1in}
\caption{\label{fig:AdityaFit500}GP fit for 500 points of level=2 (a) and level=3 (b) Aditya time series data.}
\end{figure}

Fig. \ref{fig:AdityaFit500} (a) and (b) show the fit obtained by the GP solutions (Eq. \ref{eq:GPSoln_Aditya}) for level=2 and level=3 resepctively. Both the fits are found to be quite good.

Having obtained the map equations, it is interesting to see how well these solutions work in different regions of the time series out side the fitted region. We have selected such data sets at 7 different regions beginning at data point lying between 1000 and 7000.

\begin{figure}
\centering
  \begin{tabular}{cc}
    {\resizebox{!}{3.4cm}{%
       \includegraphics{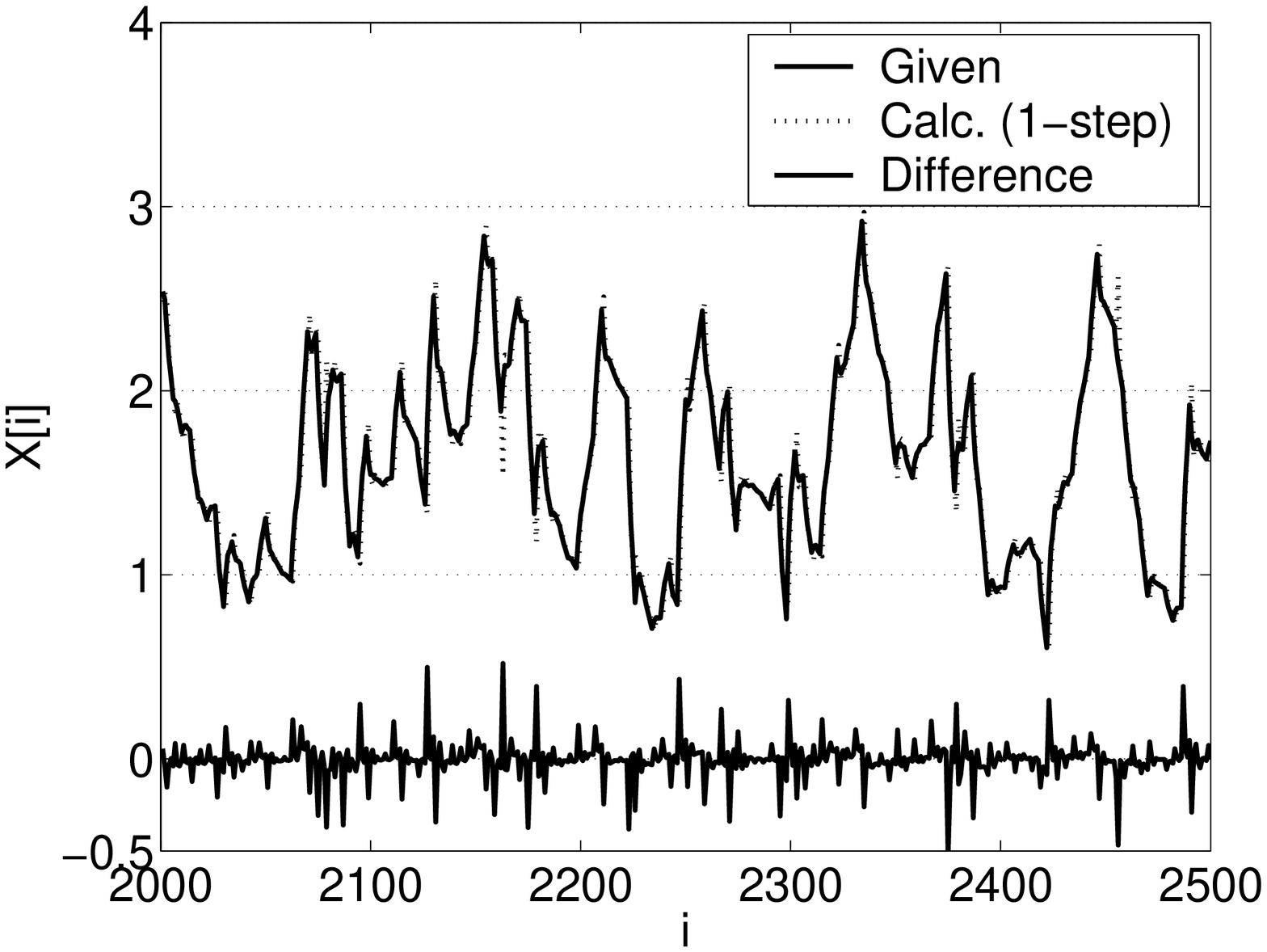}}}
    {\resizebox{!}{3.4cm}{%
       \includegraphics{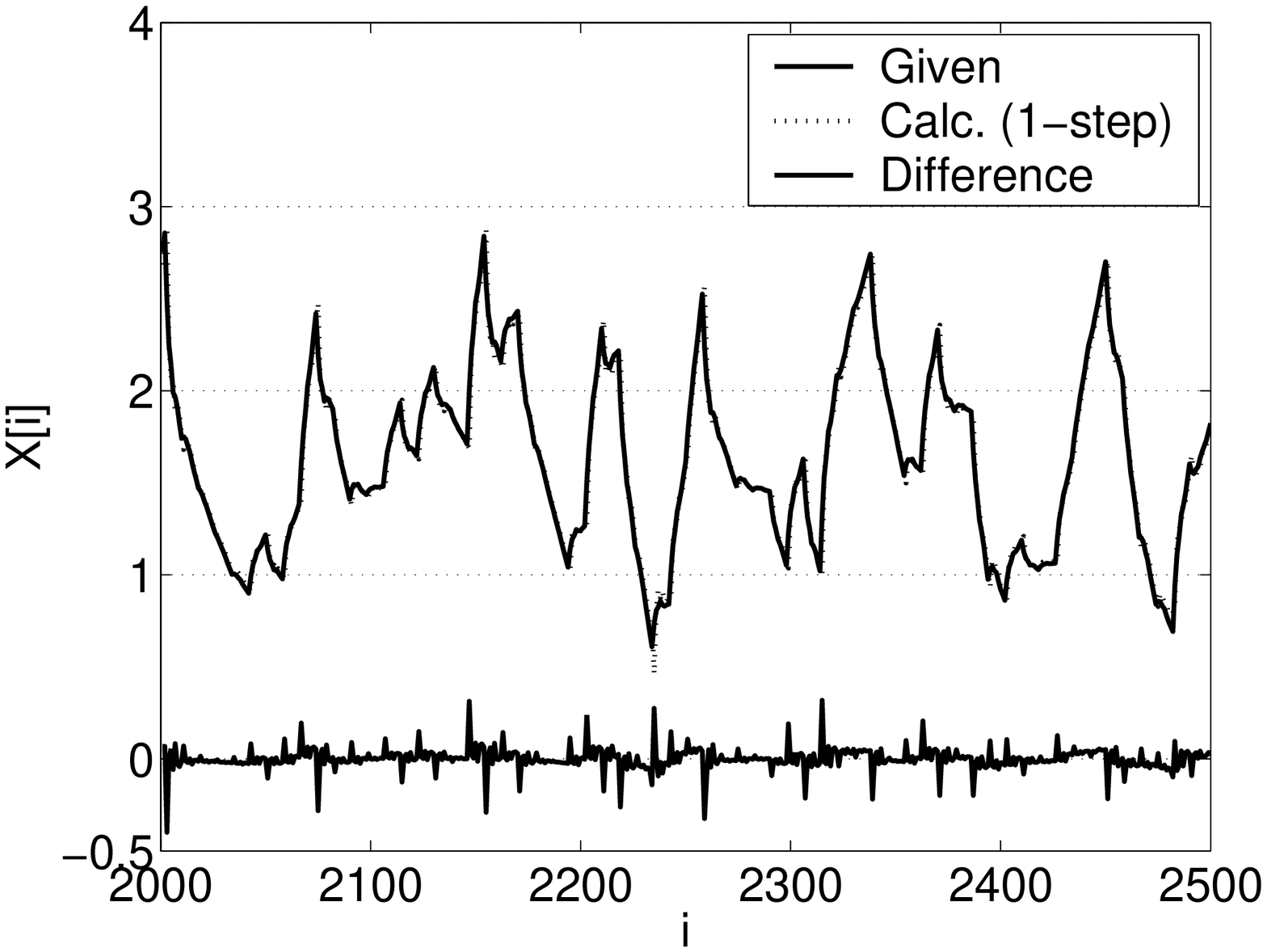}}}
  \end{tabular}
  \vspace{-0.15in}
    \begin{verbatim}              (a)                   (b)\end{verbatim}
\caption{\label{fig:AdityaFit2000}GP predictions for 500 points of level=2 (a) and level=3 (b) Aditya time series data outside the fitted region of 500 points beginning at point 2000 each.}
\end{figure}

Fig. \ref{fig:AdityaFit2000} show 1-step out of sample prediction for 500 points beginning at point 2000 for level=2 series (a) and level=3 series (b). Table \ref{tab:AdityaNMSE} shows the NMSE values for the above 7 regions in which the fit is tested. It can be seen that the 1-step prediction is on the whole quite good.

\begin{table}
\caption{\label{tab:AdityaNMSE}NMSE (Eq. \ref{eqn:NMSE}) for out-of-sample prediction for 500 points of Aditya Tokamak time series at 7 different regions lying within data point 1000 to 7000.}
\begin{ruledtabular}
\begin{tabular}{ccccc}
              &\multicolumn{2}{c}{$level=2$}&\multicolumn{2}{c}{$level=3$}        \\
Starting point&NMSE&Variance  &  NMSE&Variance       \\                   \hline
          1000&0.034&0.229    &0.019&0.217           \\
          2000&0.040&0.255    &0.017&0.233           \\
          3000&0.056&0.188    &0.032&0.174           \\
          4000&0.066&0.250    &0.025&0.208           \\
          5000&0.058&0.209    &0.031&0.187           \\
          6000&0.041&0.230    &0.026&0.215           \\
          7000&0.031&0.308    &0.023&0.278           \\
\end{tabular}
\end{ruledtabular}
\end{table}

The Lyapunov exponents for the Db-4 level=2 and level=3 Aditya time series conisdered above are 0.276 and 0.361 respectively. Using these Lyapunov exponent values, it is estimated that the calculated initial error of 0.01766 (in $501^{st}$ step) would grow in 10 steps to 0.2117 for level=2 series and the calculated initial error of 0.03779 (in $501^{st}$ step) would grow in 10 steps to 0.9737 for level=3 series. It is therefore expected that the dynamic predictions for these series would be in general difficult. However we have not made such predictions because they are not physically interesting.

\subsection{NASDAQ composite index time series}
The NASDAQ time series considered corresponds to daily closing values of the composite index for the duration 2-Mar-1998 to 27-Mar-2002. We have divided each data value of NASDAQ series by a factor of 1000 for computational convenience. This series is first smoothened with Db-4 transform with level=1, 2 and 3.

\begin{table}
\caption{\label{tab:NASDAQTauDim}Time lag $\tau$ and dimension d for Db-4 transformed level 1, 2 and 3 NASDAQ time series.}
\begin{ruledtabular}
\begin{tabular}{ccc}
 &Time Lag $\tau$&Dimension d        \\ \hline
Original NASDAQ series&2&4           \\
NASDAQ series (Db-4 level=1)&2&3     \\
NASDAQ series (Db-4 level=2)&3&4     \\
NASDAQ series (Db-4 level=3)&3&5     \\
\end{tabular}
\end{ruledtabular}
\end{table}

The time lag $\tau$ and dimension d of the embedded vectors are shown in Table \ref{tab:NASDAQTauDim}.

In the following we present results for GP solutions for Db-4 level=2 and level=3 smoothened time series. A GP fit is separately made on the datasets of these two NASDAQ time series having 500 points each. 

The GP solutions for the above two smoothened time series are shown in Eq. (\ref{eq:GPSoln_NASDAQ}) where we use the notation $X_{tm}$=$X_{t-m*\tau}$ and show double precision numbers to only 3 significant digits for the sake of simplicity.
\begin{widetext}
\begin{eqnarray}
X_{t}^{level=2}&=&\frac{5.5 X_{t1}^{2}(X_{t1}-4.305)}{7.245 X_{t1}^{2}-31.193X_{t1}+X_{t1}X_{t2}-4.305X_{t2}} \nonumber \\
X_{t}^{level=3}&=& 0.357X_{t1}+\frac{0.179X_{t2}X_{t4}(X_{t1}-X_{t2})}{X_{t5}[3.5+1.01X_{t2}+\frac{0.81(\frac{(X_{t1}-8)X_{t5}}{X_{t3}-5.6})+0.304}{5.6X_{t3}(X_{t2}-4X_{t5}+15.1)-3.7X_{t5}}+\frac{0.72}{X_{t5}[3.287+\frac{9.52}{X_{t4}(2.1-X_{t5})}]}]}
\label{eq:GPSoln_NASDAQ}
\end{eqnarray}
\end{widetext}

\begin{table}
\caption{\label{tab:NASDAQFitness}The fitness parameters for Db-4 smoothened level 2 and 3 NASDAQ time series obtained using GP fit on datasets of 500 points each.}
\begin{ruledtabular}
\begin{tabular}{ccc}
 &$\bigtriangleup^{2}$&Fitness                  \\                    \hline
NASDAQ series (Db-4 level=2)&1.02666&0.9959     \\
NASDAQ series (Db-4 level=3)&0.324&0.9986       \\
\end{tabular}
\end{ruledtabular}
\end{table}

From Table \ref{tab:NASDAQFitness}, we see that the fitness values for the two GP solutions are very good. It must be pointed out that we have enforced persistent solutions in the initial pool of chromosomes having fitness values 0.9924 and 0.9962 for level 2 and 3 series respectively to obtain the GP solutions (Eq. \ref{eq:GPSoln_NASDAQ}). This feature of enforcing persistent solutions is explained in Appendix Sec. 2.

\begin{figure}
\centering
  \begin{tabular}{cc}
    {\resizebox{!}{3.4cm}{%
       \includegraphics{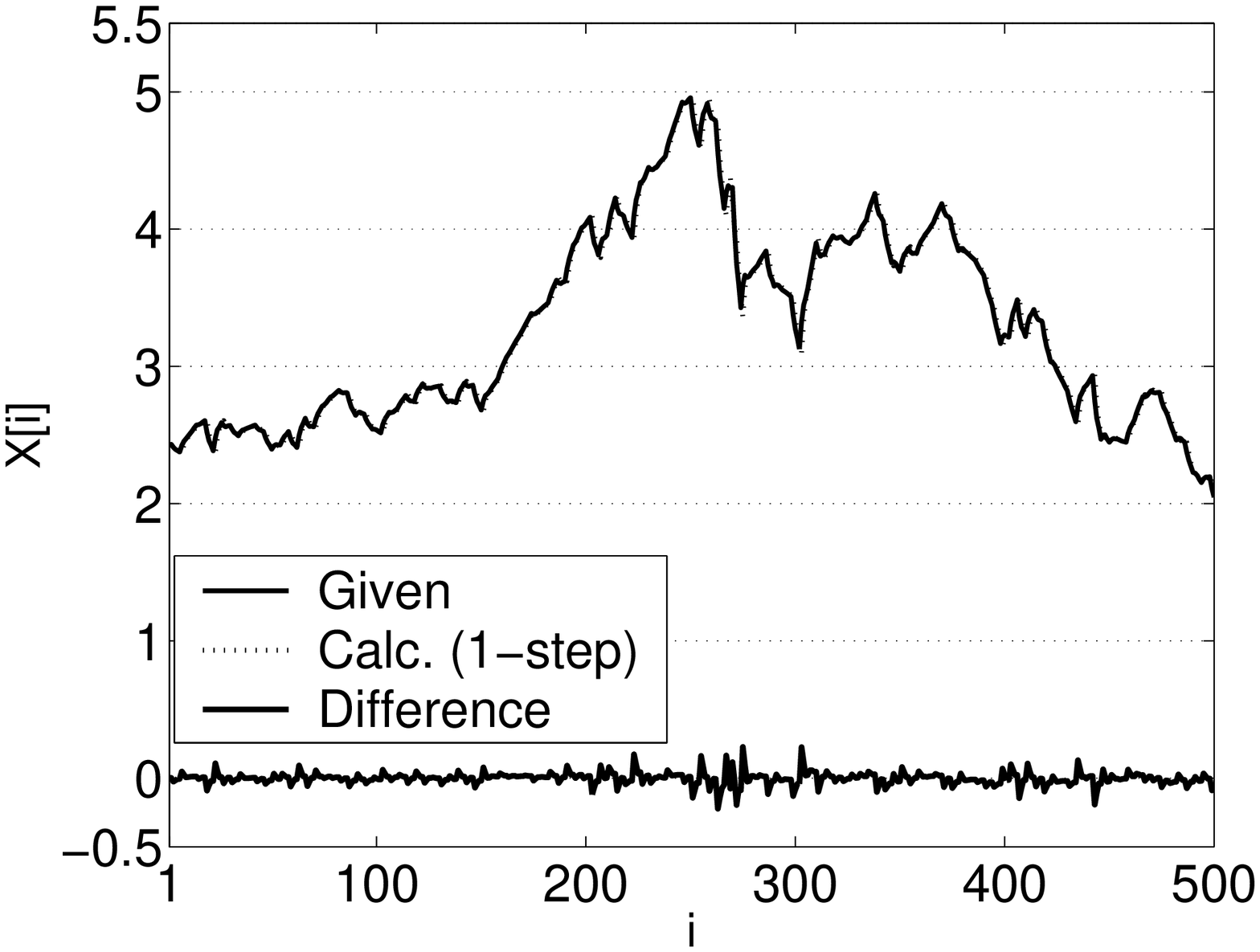}}}
    {\resizebox{!}{3.4cm}{%
       \includegraphics{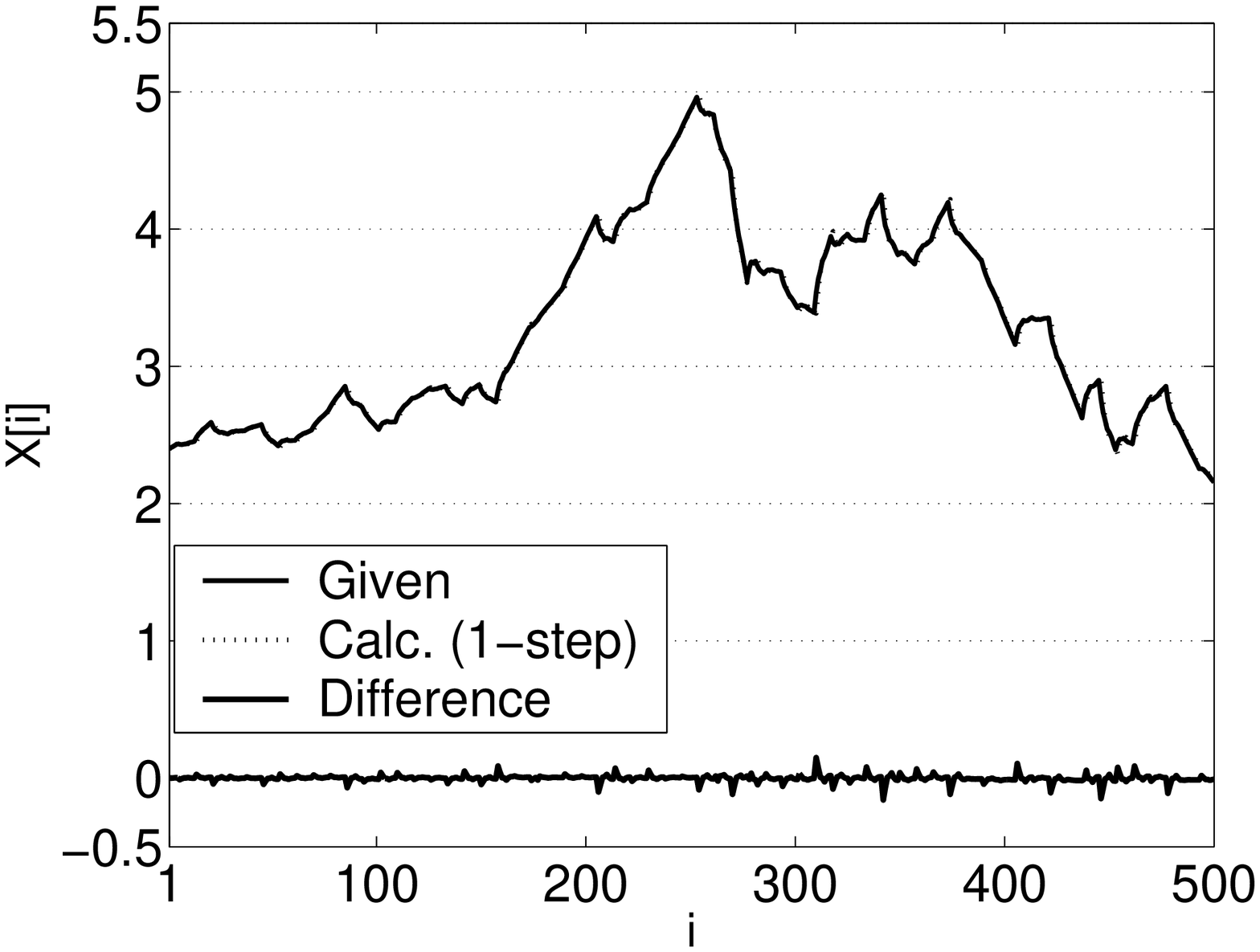}}}
  \end{tabular}
    \begin{verbatim}             (a)                        (b)\end{verbatim}
\caption{\label{fig:NASDAQ_Fit500}GP fit for datasets of 500 points for NASDAQ time series for  level=2 (a) and level=3 (b).}
\end{figure}

Fig. \ref{fig:NASDAQ_Fit500} (a) and (b) show comparision between given and GP calculated values of datasets of 500 points for level=2 and level=3 series respectively. The thick lines close to 0.0 in the figures indicate the difference between the given and calculated values and the small values for differences indicate that the fits are reasonable. The fit is better for level=3 series as compared to level=2 series.

Next we carry out a one-step out-of-sample prediction beyond the fitted dataset of 500 points.

\begin{figure}
\centering
  \begin{tabular}{cc}
    {\resizebox{!}{3.3cm}{%
       \includegraphics{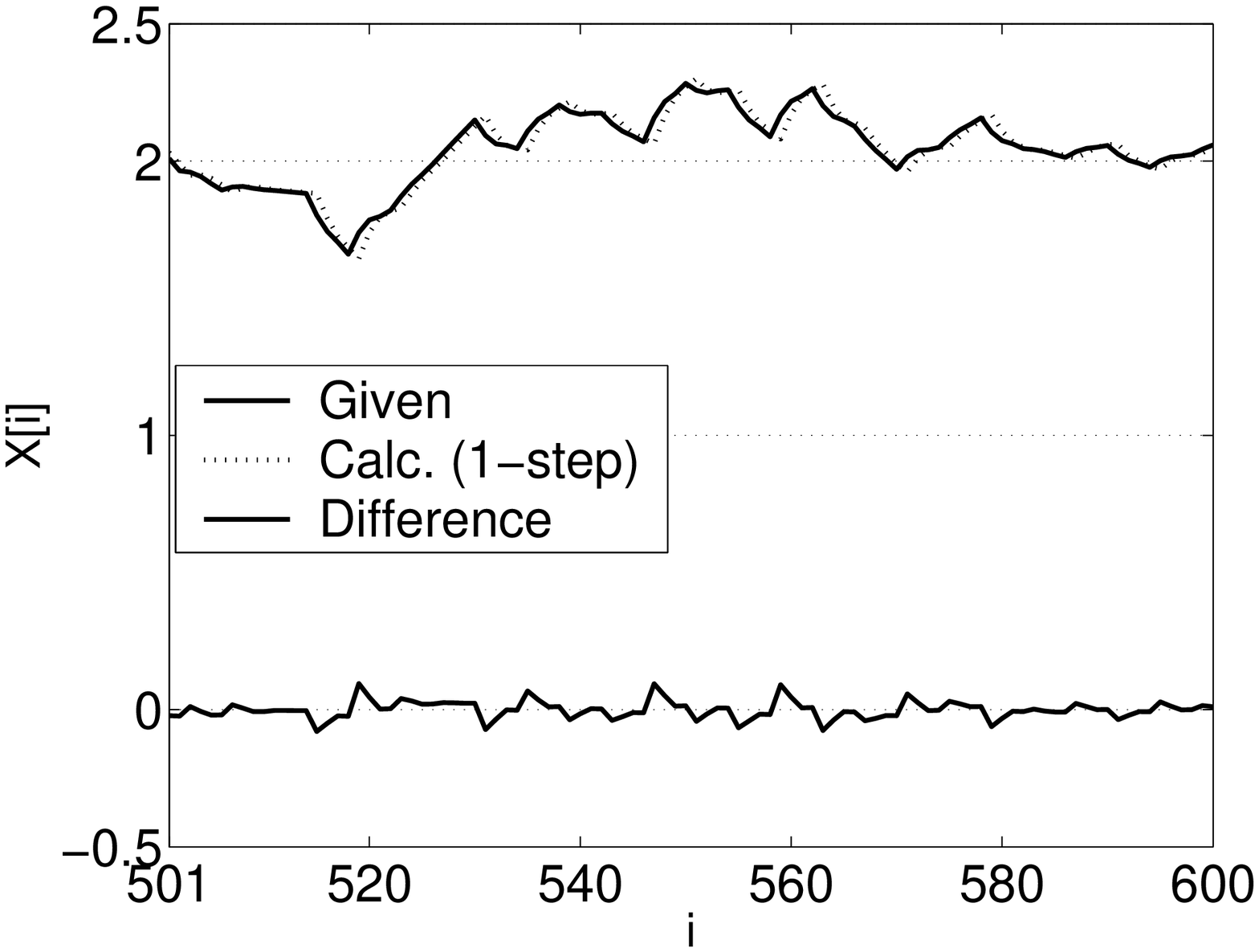}}}
    {\resizebox{!}{3.3cm}{%
       \includegraphics{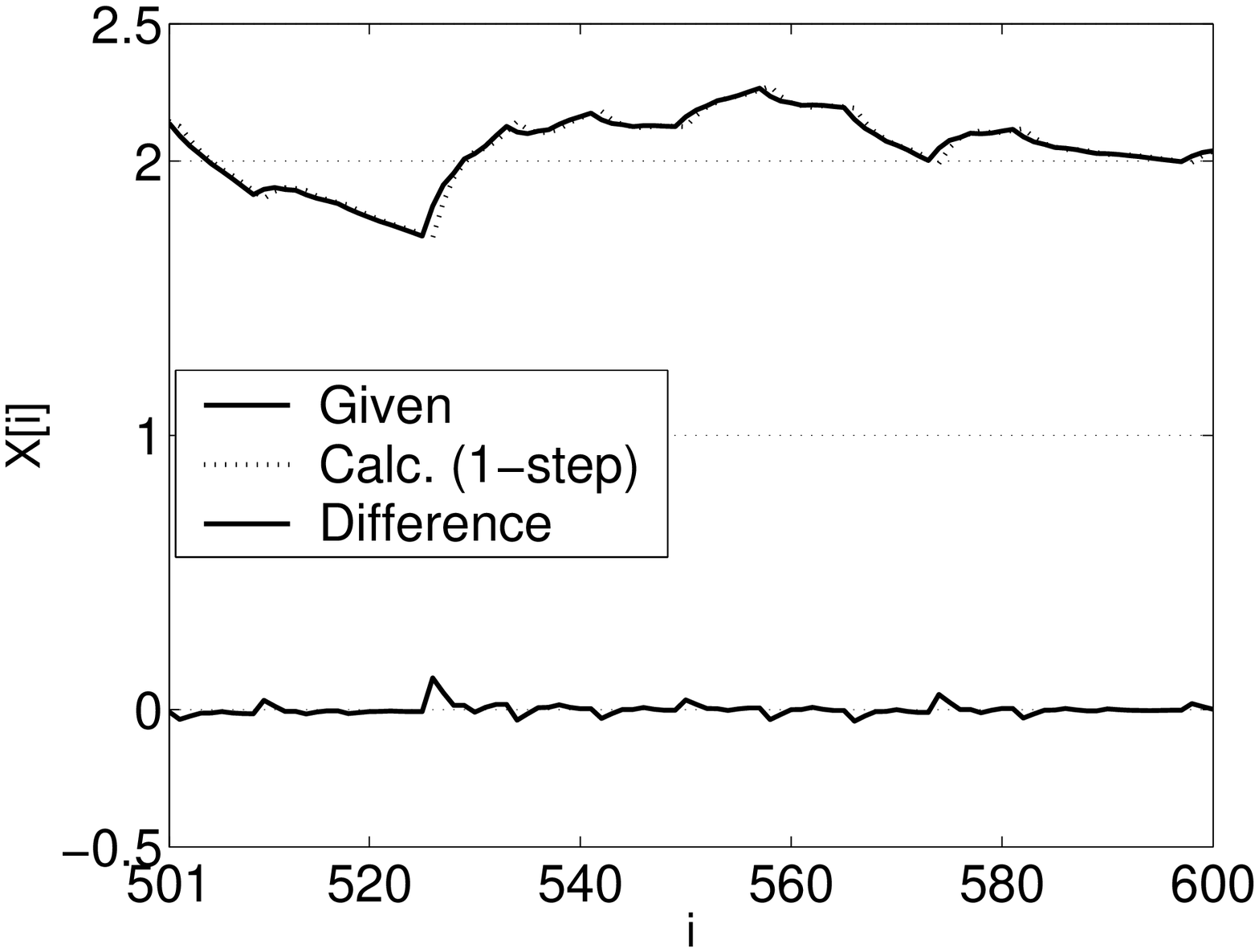}}}
  \end{tabular}
    \begin{verbatim}             (a)                        (b)\end{verbatim}
\caption{\label{fig:NASDAQ_Predict501_1step}Out of sample prediction beyond the fitted dataset of 500 points for (a) NASDAQ Db-4 level=2 time series (d=4, $\tau$=3) and (b) NASDAQ Db-4 level=3 time series (d=5, $\tau$=3)}
\end{figure}

Fig. \ref{fig:NASDAQ_Predict501_1step} (a) and (b) show one-step prediction for 500 points beginning at data point 501 for NASDAQ level=2 and level=3 smoothened time series respectively. It can be seen that the one-step prediction is quite good. The NMSE values (Eq. \ref{eqn:NMSE}) for the one-step prediction for 100 points are 0.0593 and 0.0242 for level=2 and level=3 smoothened time series resepctively.

\begin{figure}
\centering
  \begin{tabular}{cc}
    {\resizebox{!}{3.3cm}{%
       \includegraphics{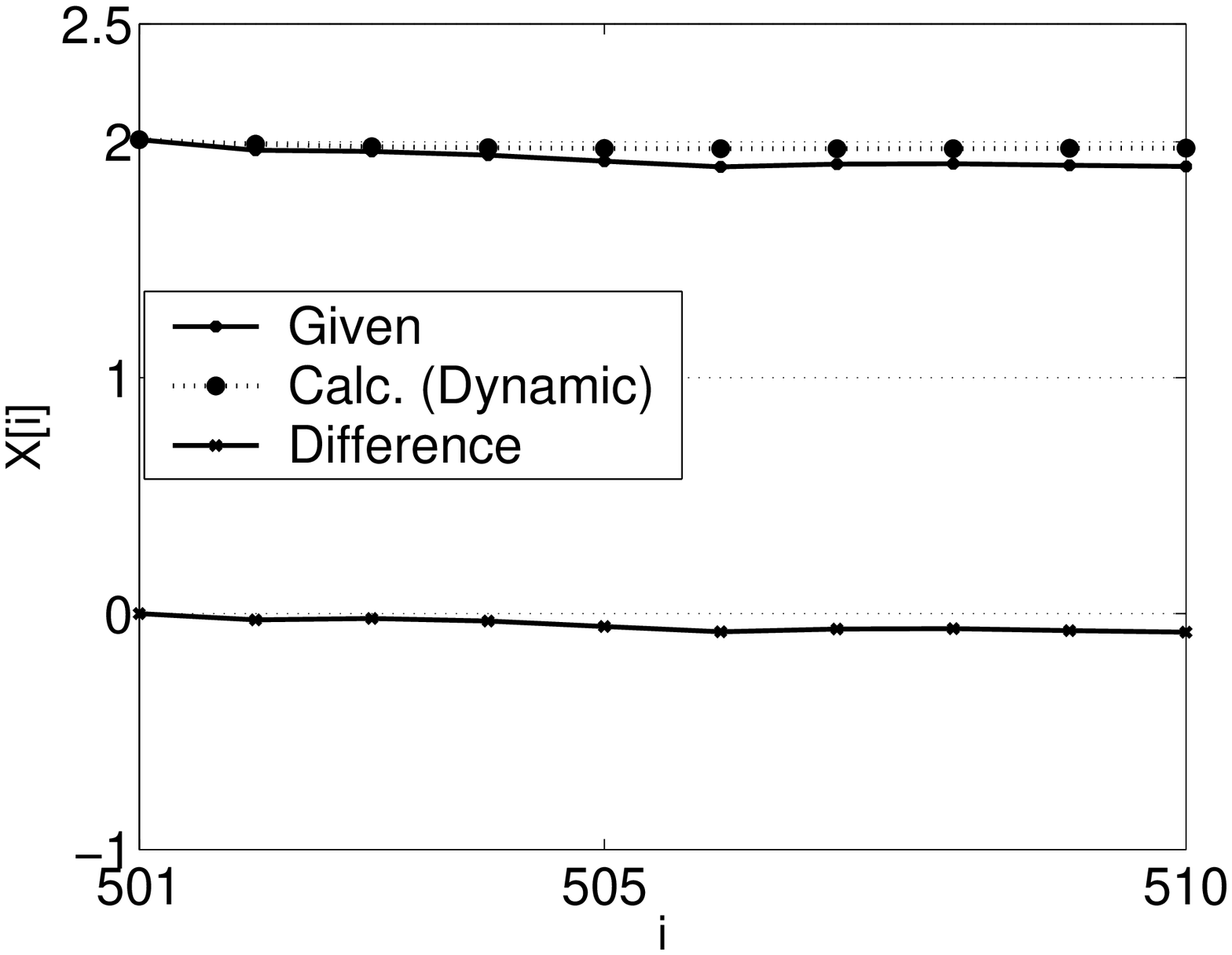}}}
    {\resizebox{!}{3.3cm}{%
       \includegraphics{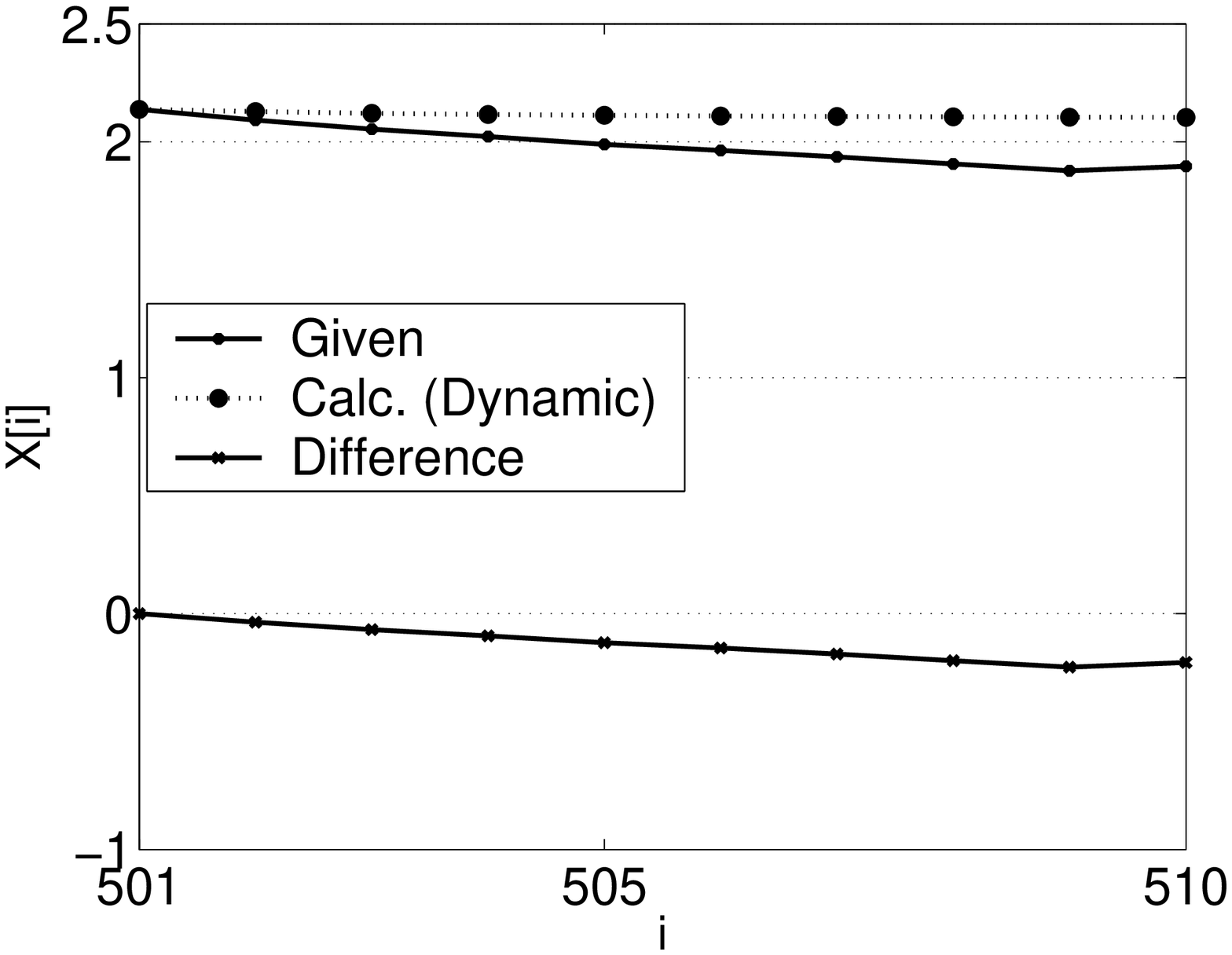}}}
  \end{tabular}
    \begin{verbatim}             (a)                        (b)\end{verbatim}
\caption{\label{fig:NASDAQ_Predict501_Mstep}Out-of-sample dynamic prediction beyond the fitted dataset of 500 points for (a) NASDAQ Db-4 level=2 time series (d=4, $\tau$=3) and (b) NASDAQ Db-4 level=3 time series (d=5, $\tau$=3)}
\end{figure}

Next a dynamic prediction is made using the GP solutions. It may be noted that the first point (i.e. data point 501) is not guaranteed to match exactly. However considering the fact that the general dynamics of the time series would have been captured by GP, the calculated series is shifted such that the data point 501 is matched with the given value. This requires a marginal shifting for the two calculated series value by -0.0219 (for level=2 series) and by -0.0073 (for level=3 series).

Fig. \ref{fig:NASDAQ_Predict501_Mstep} (a) and (b) show comparison between actual and predicted values for NASDAQ level=2 and level=3 smoothened series resepectively. As can be seen from the figure, the predictions are not good . The dynamic prediction for 10 points gives NMSE value as 2.325 and 3.102 for level=2 and level=3 series respectively.

The Lyapunov exponents for Db-4 level=2 and level=3 NASDAQ time series are 0.127 and 0.133 respectively. Using these Lyapunov exponents it is estimated that an initial error of 0.026 in $502^{nd}$ step (level=2) would grow in 10 steps to 0.093 and an initial error of 0.036 in $502^{nd}$ step (level=3) would grow in 10 steps to 0.14. In Fig. \ref{fig:NASDAQ_Predict501_Mstep} the error between calculated (dynamic) and given data values on the $10^{th}$ step are 0.078 (for level=2) and 0.21 (for level=3).

\section{Concluding remarks}
Our findings with different data sets suggest that modeling deterministic time series using Genetic Programming is a very promising approach. This is particularly the case for data of real systems having complex dynamics after the statistical fluctuations are filtered out. The filtered series is to modeled by GP and the fluctuations by their statistical properties. Also it may be noted that as against other modeling approaches, GP does not require the calculation of any derivatives within the optimization procedure, and therefore it is well suited for the modeling of rapidly fluctuating time series.

Our results for Logistic map and Rossler discretized system are quite impressive, giving good fitness values, good NMSE values, good 1-step predictions and above all good dynamic predictions. On the other hand the GP models for time series of real systems are satisfactory for giving good fitness values.

As we have seen, a short coming of this method so far is that for real systems the iterative dynamic predictions are poor. In view of this we have developed (see Appendix Sec. 3) an extension of GP to include fitting a sequence of values of the time series. This will enable us to capture patterns in the time series data. This work will be reported separately.

\begin{acknowledgments} 
We wish to acknowledge Dr. R. Jha of Institute for Plasma Research for providing us with Aditya Tokamak data \cite{Aditya:Jha}.
\end{acknowledgments}

\appendix* 

\section{Implementation of Genetic Programming}

Here we give details of implementation and improvements that have been made in our work.

\subsection{Introduction to Genetic Programming}
A Genetic Programming (GP) considers an ensemble of chromosomes, called the population, as the starting point and then processes it from one generation to the next. A given chromosome encodes a candidate solution of the optimization problem. The fitness of the chromosome is decided by an objective function that maps the chromosome structure to a fitness value. It is assumed that highly fit chromsomes are more likely to breed offsprings for the next generation. Genetic operators, namely copy, crossover and mutation are applied to generate a new ensemble of chromsomes. As a result of this evolutionary cycle of selection, crossover and mutation, more and more suitable chromsomes for the given optimization problem emerge within the population. It is at the discretion of the user to select the top (or one of the top most) chromosome of the population at the end of a sufficient number of generations of the population.

\subsection{Implementation details of Genetic Programming}
It is required to address following issues while considering GP as a means of solving an optimization problem:

\begin{itemize}
  \item Structure of chromosome: We have used a binary tree representation of the chromosome as described below.
  \item An objective function for assessing the fitness of a chromosome: Eq. \ref{eq:RSquareModified} that uses Eq. \ref{eq:SumSqErrors} and Eq. \ref{eq:RSquare} is used as an objective function.
  \item Values of parameters of the GP, such as the following:
    \begin{enumerate} 
      \item Population size: We have used 400 chromosomes for 1-step fit. 
      \item Number of generations: Maximum number of generations is set at 5000, however if the fitness of the top chromsome does not vary sizably for around a few hundred generations, then iterations are stopped.
      \item Probabilities for appying genetic operators: Whenever an operator needs to be chosen at random, we use 40\% and 60\% for selecting either a number (from -10.0 to 10.0 with the precision of one decimal) or a time lagged variable. For the mutation probability we have used 90\% for enforcing mutation leaving aside top 10\% chromosomes from getting mutated. This is referred to as {\it elitism} in GA literature that helps preserve a proportion of top most population in successive generations thereby not loosing whatever good that has been found so far.
    \end{enumerate}
\end{itemize}

For the optimization problem being considered presently, the structure of a chromosome is an equation of the following form:

$X_{t} = f(X_{t-\tau}, X_{t-2\tau}, X_{t-3\tau}, ... X_{t-d\tau})$

where d is the embedding dimension and $\tau$ is the sampling time. The aim of Genetic Programming is to find out the best possible functional form of f that gives rise to the map of the given time series in terms of the previous time lagged components. Once the map equation is available by fitting a given set of data points, it can be used for predicting the future state of the system, i.e. generate out-of-sample time series data. 

The structure of chromosome is in the form of an algebraic expression involving binary operators, numbers and time lagged variables of the form $X_{t-\tau}$. We have used a binary tree representation using non-linear dynamic data structures for the structural representation of chromosomes.

The binary tree structure has several advantages over a linear structure of characters: 
\begin{itemize}
  \item Brackets are not required to be stored explicitly in the binary representation.
  \item The genetic operations on the chromosomes, are considerably simplified as only pointers need to be manipulated while carrying out the crossover operation.
  \item Since binary tree is grown using a dynamic data structures, it is not required to specify an upper limit for number of tokens in the expression tree. This eliminates the need for boundary level checking for the overflow of the size of expression tree beyond the specified limit. In such a case of overflow of expression size, one normally brings back the expression to a template structure thereby loosing the structural information found so far.
  \item It has been found that many of the GP solutions turn out to be in the Pade form. It is quite straight forward to check whether a given GP solution is in the Pade form or not. This is done just by inspecting the operator in the root node of the tree and confirming whether it is division operator or not. In the same way, it is quite straight forward to optionally impose a Pade form for GP solution.
  \item Evaluation of the expression in binary tree representation is considerably simpler (each leaf sub-tree is evaluated and collapsed to a numerical value using a recursive procedure till the whole tree finally reduces to a number). In contrast, if an expression has a linear structure (e.g. an array), then multiple passes of array are required to give due credence to the hierarchy of operators.
\end{itemize}

It is observed that for a non-stationary time series (e.g. NASDAQ series), a persistent solution $X_{t}$=$X_{t-1}$, although not the best possible solution, usually gives quite a good fitness value close to 1.0. Further it is observed that if during GP iterations such persistent solutions are found, they tend to dominate the population thereby giving rise to convergence to such trivial solutions. Since these trivial solutions are of no interest from the viewpoint of underlying dynamics of the system, it is essential to get the GP away from such trivial solutions. It is interesting to understand the reason why the majority of the population tends to get flooded by trivial solutions. This is because the tree for such solutions is a single node tree and so crossover cannot lead to new solutions. Mutation would also not help in such situations because GP settles down to a single node solution corresponding to the best possible time lagged variable (say $X_{t-1}$), and changing this variable to another time lagged variable leads to decrease in fitness value. Thus GP optimization is virtually helpless i.e. is left without any means to come out of the local minima arising out of persistent solution.

In order to resolve this problem, we use various forms of multipliers to the GP chromosome. The multiplier can be a constant number C or possibly a Gaussian curve passing through the data set of points to be fitted. In case of a constant multiplier, the trivial solution is in the form of $X_{t-1}/C$ and it is easy to see that GP can now possibly vary and grow this structure by the genetic crossover and mutation operators to optimize the fitness value. In fact, we have been able to transform the problem (arising out of convergence to local minima due to such trivial solutions) to our advantage by starting the GP iteration from the reasonably higher fitness values of trivial solution. On the other hand if we start with a purely random population, we have usually found that the top most chromosome in the population has a much lower value of fitness compared to that of a trivial solution and it has to undergo a substantially large number of iterations to reach to this fitness value if not surpass it.
  
\subsection{Multi-step GP fit}
The usual method adopted for prediction is to first fit a given data set of points and then make out-of-sample prediction using either 1-step or multi-step (dynamic) prediction. The 1-step predictions obtained by GP fit are usually found to be good. The real test of any dynamic model however lies in whether it is able to make out-of-sample dynamic predictions outside the range of data set used for fitting by GP. It is observed that the dynamic prediction is either not good or at the most it gives good predictions only for a very few number of points. In a way this is understandable because our fit itself is of 1-step nature. The number of equations to be fitted by GP during an iteration are all trying to fit the next point using a map involving its time lagged variables occurring in immediate past. Thus GP is not trained to make multiple time step predictions. We have therefore also considered a multi-step fit by GP. We require GP to fit not just the next step $X_{t}$ (of Eq. \ref{eq:MapEquation}), but also $X_{t+1}$, $X_{t+2}$, ... $X_{t+m}$, where m is a predefined number of steps. Thus for a given equation to be fitted, the sum of squared errors invloves not only $[X_{i}^{calc} - X_{i}^{given}]^{2}$, but in addition $[X_{i+1}^{calc} - X_{i+1}^{given}]^{2}$, ... up to $[X_{i+m}^{calc} - X_{i+m}^{given}]^{2}$. However it may be noted that multi-step GP fit leads to a very involved computation and we are currently able to carry out such GP fit for a limited population size only.

\end{document}